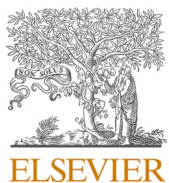

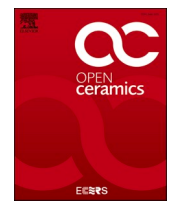

# A comparison of syntheses approaches towards functional polycrystalline silicate ceramics

Franz Kamutzki [*], Maged F. Bekheet, Sven Schneider, Aleksander Gurlo, Dorian A.H. Hanaor

*Technische Universität Berlin, Faculty III Process Sciences, Institute of Material Science and Technology, Chair of Advanced Ceramic Materials, Straße des 17. Juni 135, 10623, Berlin, Germany*

## ARTICLE INFO

## ABSTRACT

This study aims to shed light on processing pathways towards functional silicate ceramics, which show some promise in various emerging applications, including dielectrics and bioactive implant materials. Polycrystalline silicate ceramics of Neso-, Soro- and Inosilicate families were synthesised by three different techniques: (i) a co-precipitation method, (ii) a modified sol-gel method and (iii) standard solid-state reactions. Co-precipitated samples show increased sintering and densification behaviour compared to sol-gel and solid-state methods, with diametral shrinkage values during sintering of 28.8%, 13.3% and 25.0%, respectively. Well-controlled phase formation in these ceramics was most readily achieved through the steric entrapment of cations and shorter diffusion pathways afforded by the modified Pechini-type sol-gel method. Substituting $Zn^{2+}$ for $Mg^{2+}$ in enstatite samples was found to enhance the formation of orthoenstatite during cooling, which is otherwise very slow. We present guidelines for the design of synthesis methods that consider the requirements for different functional silicate ceramics in terms of phase formation and microstructure.

## 1. Introduction

Silicate ceramics are of emerging relevance as functional materials in a broad range of applications. Numerous crystalline silicate phases of neso-, soro-, ino-, cyclo- and tectosilicates, differentiated by linking of $SiO_4$ tetrahedra, have been found in recent years to exhibit promising characteristics towards applications as high-frequency dielectrics [1], bioactive ceramics [2,3], battery electrodes [4] and tritium breeders [5–8]. Silicates are promising multifunctional materials, whose utilisation hinges on our ability to fabricate such materials in controllable and appropriate forms and stoichiometries. In particular, polycrystalline silicate ceramics often suffer from poor sinterability [9–11] without the addition of sintering aids, which poses a challenge towards realising good functionality in numerous applications, particularly those that benefit from dense microstructures. A survey of reported studies reveals that manifold and often simultaneously occurring reaction pathways often frustrate the obtainment of target stoichiometries and phases [12–15]. The properties of silicate ceramics are controllable in part through cation substitutions. Extensive substitutions are commonly possible at octahedral cation sites in silicates of all families. Many such substitutions are known from the world of geosciences, and when introduced in controlled quantities in the context of ceramics processing, can be used to tune the physical and chemical behaviour of applied silicates [1]. Substitutions at tetrahedral Si sites with other cations such as Al could offer further pathways for materials design.

A vast majority of studies into functional polycrystalline silicate ceramics have employed solid state syntheses, using oxides, or occasionally carbonates, as precursors. Nevertheless, several previous studies have utilised soft-chemistry type syntheses, most prominently sol-gel and co-precipitation and, to a lesser degree, hydrothermal approaches. Generally, soft chemistry syntheses produce ceramics at lower temperatures and in forms that often diverge significantly from typical materials formed by solid state reaction of solid precursors. Nanocrystalline microstructures and non-equilibrium (i.e. metastable) phase compositions are readily observed in product materials. Harnessing the advantages offered by these methods requires new insights into how synthesis parameters can direct phase formation and microstructure.

The only comprehensive study that compares solid-state, co-precipitation and sol-gel techniques for synthesising polycrystalline silicate ceramics, focused on the formation of lithium silicates [16]. In the latter work, Pfeiffer et al. demonstrated that the choice of synthesis route has profound effects on phase formation. Specifically, the sol-gel technique investigated in that work was found to offer advantages when aiming for ceramic products with Li:Si molar ratios of 2, while solid state and precipitation methods were recommended for Li:Si molar ratios of 4. This was because some lithium silicate phases are formed by polymerisation reactions while others are produced in a co-precipitation.

---

* Corresponding author.
  *E-mail address:* franz.kamutzki@ceramics.tu-berlin.de (F. Kamutzki).

In this work, we examine several synthesis methods, including solid-state and soft-chemistry approaches, for the formation of neso-, soro- and inosilicates. We characterise materials in terms of phase formation, stoichiometry, morphology and reaction mechanisms to assess the suitability of the methods towards the control of phase assemblage and sintering. In this context, the effect of isovalent cation substitution in silicates is further explored. For this purpose, we chose $Zn^{2+}$ for $Mg^{2+}$ substitution as these two elements should easily substitute for each other according to Goldschmidt's rules [17,18], as they have the same oxidation state and their ionic radii are very similar ($Zn^{2+}$, 6-fold coordinated = 0.74 Å; $Mg^{2+}$, 6-fold coordinated = 0.72 Å [19]). Several previous works have demonstrated the $Zn^{2+}$ for $Mg^{2+}$ substitution in silicate ceramics, and in some cases, with a focus on the dielectric performance of these silicate ceramics. For instance, a positive effect on the loss characteristics has been reported for $Mg_{2-x}Zn_xSiO_4$ [20], $LiMg_{1-x}Zn_xSiO_4$ [21], $Mg_{2-x}Zn_xSi_2O_6$ [22] and $CaMg_{1-x}Zn_xSi_2O_6$ [23].

## 2. Materials and methods

### 2.1. Synthesis of ceramic powders

For this study, silicate specimens of eight compositions with varying degrees of the connectivity of [SiO₄] tetrahedra were synthesised. These eight targeted silicates contained four end-member compositions and four Zn- or Mg-substituted varieties thereof. The respective ceramic powders were prepared for each material via a standard solid-state reaction route, a Pechini type sol-gel path and a co-precipitation approach. The prepared material compositions are listed in Table 1. A detailed description of precursor quantities for each sample synthesis is tabulated in the supporting information (Table S1).

For the solid-state (SS) approach, dried high purity binary metal oxide powders of MgO (Merck, Germany, 99%), ZnO (Merck, Germany, 99%), CaCO₃ (Carl Roth, Germany, 99%) and amorphous SiO₂ (Merck, Germany, 99%) were mixed in stoichiometric ratios and given into polyoxymethylene jars with 250 ml deionized water, ball milled for 6 h with ZrO₂ milling balls and subsequently dried at 80 °C in ambient atmosphere.

A type of sol-gel method, named after its inventor Maggio Pechini (P) [24], was chosen as a soft chemistry route to achieve a homogeneous powder product on a molecular level. This type of sol-gel synthesis was chosen as the steric entrapment afforded by the chelation-polymerisation pathway is known to facilitate homogeneity in oxide systems with multiple cation sites [25,26]. A total molar ions: citric acid: ethylene glycol molar ratio of 1:6:6 was used in this work. In the synthesis of Pechini type materials here, hydrous nitrate precursors of $Mg(NO_3)_2*6H_2O$ (Merck, Germany, 99%), $Zn(NO_3)_2*6H_2O$ (Carl Roth, Germany, 99%) and $Ca(NO_3)*4H_2O$ (Carl Roth, Germany, 99%) were used. These precursors, in appropriate quantities, were added into 500 ml of a stirred aqueous citric acid solution (pH = 3) to form a homogeneous solution of metal-citrate chelate complexes. $SiC_8H_{20}O_4$ (tetraethyl orthosilicate, TEOS, Carl Roth, Germany, 99%) was subsequently added dropwise to prevent rapid precipitation. The key reaction in this Pechini method then takes place between citrate groups and the subsequently added ethylene glycol (Merck, Germany). This transesterification reaction produces a covalent polymer network via heating the solution to ~80 °C. After reduction into a thick resin, the material was heated with 3 K/min in ambient atmosphere to 400 °C and held for 6 h in a muffle furnace to remove the major part of its volatiles without initiating crystallisation. The product was then ground in an agate mortar.

Lastly, in a co-precipitation method (CP), chloride precursors of

## Table 1

Sample overview with target compositions, processing details and properties. Abbreviations not declared in the table: OE – orthoenstatite, PE – protoenstatite, Di – diopside, Ps – pseudowollastonite, Ak – akermanite, Mr – merwinite, Mo – monticellite. The open porosity and density were measured for the ceramic pellets after the sintering process. All chemicals used are listed in the supporting information (Table S1). Quantification of all sintered pellet's crystalline phase assemblage is given in Fig. 1.

| Sample | | Synthesis method | Phase composition | |
|---|---|---|---|---|
| Targeted composition and phase (silicate group) | Sample, calcination/ sintering temperature | Co-precipitation (CP) | Pechini (P) | Solid state (SS) |
| $Mg_2SiO_4$, *forsterite (Fo, nesosilicate)* | Calcined powder, 1100 °C | OE, Fo | MgO, Fo, OE | MgO, Fo, OE |
| | Pellet, 1400 °C | CE, Fo | Fo, $SiO_2$, MgO | Fo, MgO, $SiO_2$ |
| | Open porosity, % | did not sinter | 40.81 | 2.15 |
| | Density [g/cm³] (% theoretical) | | 1.81 (56.1) | 2.97 (92.1%) |
| $Mg_{1.8}Zn_{0.2}SiO_4$, *forsterite (Fo, nesosilicate)* | Calcined powder, 1100 °C | Fo, OE | MgO, Fo, OE | MgO, Fo, OE |
| | Pellet, 1400 °C | Fo, CE | Fo, MgO | Fo, MgO, $SiO_2$ |
| | Open porosity, % | 0.94 | 32.92 | 0.98 |
| | Density [g/cm³] (% theoretical) | 3.28 (94.7) | 2.15 (62.0) | 3.30 (95.3) |
| $Zn_2SiO_4$, *willemite (Wi, nesosilicate)* | Calcined powder, 1100 °C | Wi, ZnO | Wi, ZnO | Wi, ZnO |
| | Pellet, 1350 °C | Wi | Wi, ZnO | Wi, ZnO |
| | Open porosity, % | 0.08 | 16.85 | 0.23 |
| | Density [g/cm³] (% theoretical) | 4.07 (95.8) | 3.26 (76.6) | 4.07 (95.8) |
| $Zn_{1.8}Mg_{0.2}SiO_4$, *willemite (Wi, nesosilicate)* | Calcined powder, 1100 °C | Wi, ZnO | Wi, ZnO | Wi, ZnO |
| | Pellet, 1350 °C | Wi | Wi, ZnO | Wi, ZnO |
| | Open porosity, % | 0.18 | 0.25 | 0.34 |
| | Density [g/cm³] (% theoretical) | 3.92 (95.5) | 3.81 (93.0) | 3.99 (97.3) |
| $Mg_2Si_2O_6$, *clinoenstatite (CE, inosilicate)* | Calcined powder, 1100 °C | OE | MgO, Fo, OE | Fo, OE |
| | Pellet, 1450 °C | PE, $SiO_2$, CE | CE, Fo | CE, Fo |
| | Open porosity, % | 0.10 | did not sinter | 3.17 |
| | Density [g/cm³] (% theoretical) | 2.76 (86.1) | – | 2.89 (90.2) |
| $Mg_{1.8}Zn_{0.2}Si_2O_6$, *clinoenstatite (CE, inosilicate)* | Calcined powder, 1100 °C | OE | ZnO, MgO, Fo, OE | Fo, OE |
| | Pellet, 1450 °C | OE, CE, $SiO_2$ | CE | CE, Fo, OE, ZnO |
| | Open porosity, % | 5.44 | 22.34 | 2.26 |
| | Density [g/cm³] (% theoretical) | 2.97 (89.0) | 2.42 (72.8) | 2.90 (87.0) |
| | | Di, Ps | Ak, CaO, Di, Mr | Mr, Ak |

*(continued on next page)*

**Table 1** (*continued*)

| Sample | Synthesis method | Phase composition | | |
|---|---|---|---|---|
| Ca2MgSi2O7, *akermanite (Ak, inosilicate)* | Calcined powder, 1100 °C | | | |
| | Pellet, 1300 °C | Di, SiO2 | Ak, Mo, Mr | Mr, Mo, Ak |
| | Open porosity, % | 1.09 | 19.56 | 22.45 |
| | Density [g/cm³] (% theoretical) | 1.99 (66.7) | 2.14 (71.6) | 2.82 (94.3) |
| Ca2Mg0.9Zn0.1Si2O7, *akermanite (Ak, inosilicate)* | Calcined powder, 1100 °C | Di | Ak, CaO, Di, Mr | Mr, Ak |
| | Pellet, 1300 °C | Di, SiO2 | Ak, Mr, Mo | Mr, Mo, Ak |
| | Open porosity, % | 27.05 | 22.56 | 5.29 |
| | Density [g/cm³] (% theoretical) | 1.99 (66.7) | 2.14 (71.6) | 2.82 (94.3) |

$MgCl_2 \cdot 6H_2O$ (Carl Roth, Germany, 99%), $ZnCl_2$ (Carl Roth, Germany, 97%), $CaCl_2$ (Carl Roth, Germany, 98%) and TEOS (Carl Roth, Germany, 99%) were dissolved in stoichiometric ratios in 600 ml ethanol. After stirring for 1 h, 30 ml of 25% $NH_3$ solution (Carl Roth, Germany) was added dropwise to adjust the pH value of the solution to ~10 and initiate precipitation. The suspension was stirred overnight and then stopped to decant the supernatant from the precipitate. The product was washed using deionized water, followed by centrifuging and decanting. This washing process was repeated three times, followed by drying the powder at 80 °C in air atmosphere.

All obtained powder precursors from the three approaches after the drying step were heated with 3 K/min and calcined in air atmosphere at 1100 °C for 4 h, then ball-milled for 6 h in a polyoxymethylene jar with 250 ml deionized and $ZrO_2$ milling balls and dried at 80 °C in ambient atmosphere.

### 2.2. Sintering of ceramic pellets

To further explore phase formation and microstructure in silicate ceramics, calcined powders were compacted and sintered as follows. Cylindrical pellets of 20 mm diameter were first uniaxially pressed (30 Mpa) using a 5% polyvinyl alcohol (Merck, Mw approx. 60000) solution as a pressing aid. To facilitate improved compaction, and thus fully explore the phase formation ability of these materials and its dependence on the synthesis route, the pellets were further pressed cold isostatically (Hofer Hochdrucktechnik, Germany) with a pressure of 2 kbar, which reduced the porosity of the green body. Sintering temperatures for each composition were chosen on the basis of reported and observed behaviour for these phases to allow good levels of densification and reaction of intermediate phases without the risk of incipient melting and consequent phase inhomogeneity in product materials. The pellets were heated first at 600 °C for 6 h to combust excess polyvinyl alcohol, followed by sintering at 1300–1450 °C for 6 h in ambient atmosphere. A heating rate of 5 K/min and furnace cooling was used.

### 2.3. Characterisation

Crystalline phase analysis was carried out on powder samples with a D8 Advance X-ray diffractometer in Bragg-Brentano geometry (Bruker, USA), equipped with a LynxEye detector and Cu radiation (λ = 1.5406 Å). Rietveld refinement of XRD patterns was performed using the FULLPROF software [27]. The resolution parameters of the diffractometer were determined from the Rietveld refinement of XRD data for $LaB_6$ standard. For bulk phase identification, the sintered pellets were ground thoroughly in an agate mortar.

Particle size distributions of uncalcined (SS and CP) and calcined (SS, CP and P) samples were determined in de-ionized water-dispersed form using the Universal liquid module of a LS13320 laser diffraction particle size analyser (Beckman Coulter, USA). Due to its gel form, particle sizes could not be obtained for unfired P materials.

Microstructural and element distribution investigations were conducted using a Gemini Leo 1530 scanning electron microscope (Carl Zeiss Microscopy GmbH, Germany) with acceleration voltage ranging from 3 to 10 keV. To reveal grain boundaries on sintered samples, the pellets were polished and thermally etched for 30 min at 100 °C below the respective sintering temperatures. Powder samples were directly attached to sample holders using carbon pads, while pellet samples were additionally fixed with Cu-tape.

The sintered pellets' density and open porosity values were determined by Archimedes' method, according to ASTM C-373-18 [28].

Thermogravimetric and differential thermal analysis data were obtained from heating in a synthetic air atmosphere (80% $N_2$, 20% $O_2$) with a STA 409 PC (Netzsch, Germany). Firstly, precipitated and dried co-precipitated powder, unfired Pechini gel mixed and ball milled solid-state derived powders of target composition $Mg_2SiO_4$ were analysed in the temperature range 20–800 °C. Secondly, calcined (1100 °C) powders of the same composition were analysed in the range 20–1400 °C.

To glean further insights into the sintering behaviour of the samples and its dependence on synthesis methods, side-view hot stage microscopy experiments (Hesse instruments, Germany) were carried out on 3 × 3 mm size cylinders produced by hand pressing powders (calcined and milled samples of $Zn_2SiO_4$ CP, P and SS with nominal composition $Zn_2SiO_4$) with 1.5 N/mm². The measurements were conducted twice for each sample (with similar results) with a temperature profile of heating rate of 5 K/min to 1350 °C, followed by holding this temperature for 30 min. During this measurement technique, the shadow of the cylindrical sample is recorded by a digital image processing system that detects geometry changes during heating. This way, critical temperatures during, e.g. the sintering of a sample, can be derived. A detailed description of the method is given elsewhere [29].

## 3. Results

### 3.1. Phase composition

Phase assemblages extracted from XRD analysis for calcined powders as well as sintered pellets are shown in Table 1. For sintered pellets, the crystalline phases have been additionally quantified by Rietveld refinement and are presented graphically in Fig. 1. Full conversion of precursor powders into the desired composition is almost completed for $Zn_2SiO_4$ by all three routes after the calcination process at 1100 °C. For most compositions, however, this calcination temperature typically yielded powders with low crystallinity and various intermediate phases, as is generally the case for the synthesis of silicates.

From Fig. 1, it is clear that sintered samples produced by the precipitation route only yielded the desired phase-pure composition for pristine and magnesium-substituted willemite. For most sintered materials, target phases were either only present in smaller quantities (precipitated samples with forsterite and enstatite target composition), and for the case of precipitated materials with a target akermanite composition, the targeted phase was entirely absent. Across all compositions, Pechini-type syntheses yielded the best results in terms of achieving the desired phases in sintered pellets. Only in the synthesis of akermanite and unsubstituted enstatite was the Pechini method unable to produce near-phase-pure materials after sintering. In the case of materials prepared by the solid state method, target phases are always present in the sintered samples, however, in varying amounts. Solid state synthesis was found to be quite effective in yielding target orthosilicate phases while the formation of pyroxene and sorosilicates (enstatite and akermanite phases), formation kinetics in the solid state

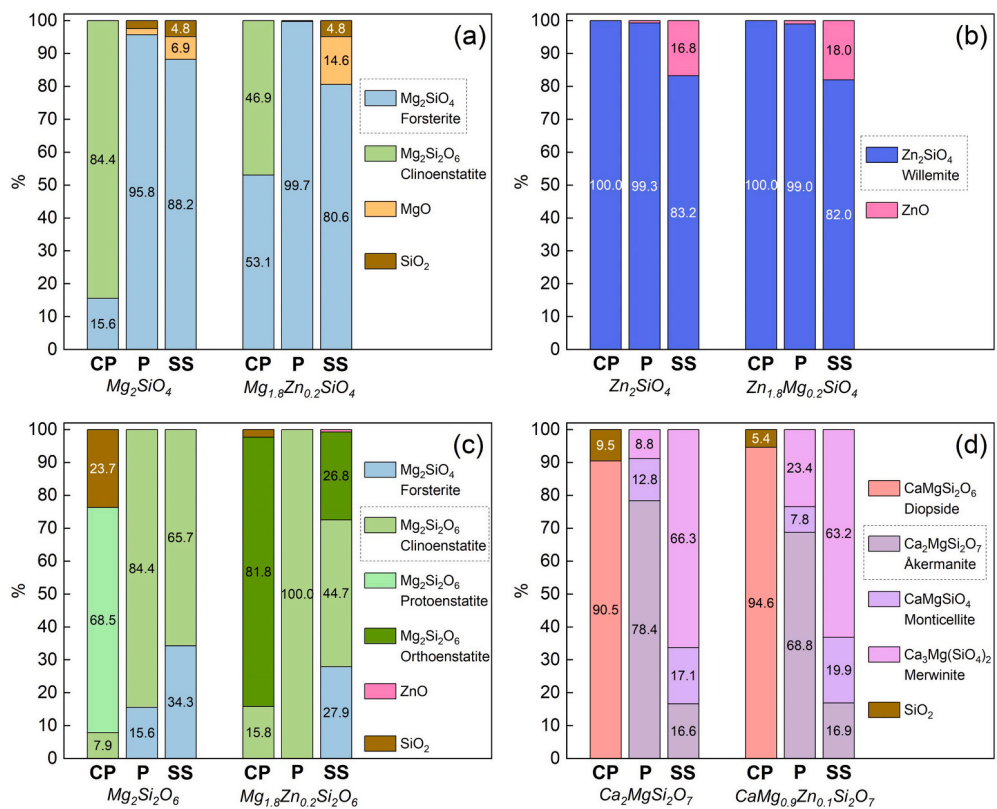

**Fig. 1.** Quantified crystalline phase compositions (wt%) for sintered pellets, with target phases framed in dashed lines. Nominal compositions for constituent phases are given in percentages, indicating major, minor or trace amounts. The SiO2 phase quantified here is predominantly cristobalite. All values are obtained by quantification of the crystalline phases in the XRD patterns with Rietveld refinement. All patterns and exact values can be found in the supporting information (Table S2 and Fig. S2).

method appear to be rather sluggish.

By comparing the phase composition of the calcined powders with that in the sintered specimens, we can get an indication of intermediate phases formed and their tendency to convert into the target silicate phases during the sintering step. Samples calcined at 1100 °C contain extensive amorphous silica, which reacts further in the subsequent sintering step, for example resulting in the further conversion of ZnO to willemite. This is particularly true for materials formed by co-precipitation, which exhibit lower crystallinity following calcination as indicated by a higher amorphous halo signal at low 2θ angles (i.e., between 20° and 30°) of diffraction patterns, indicating short-range ordered Si–O. For the calcium-containing materials ($Ca_2MgSi_2O_7$ and $Ca_2Mg_{0.9}Zn_{0.1}Si_2O_7$) prepared by CP, the phase assemblage was dominated by diopside after sintering, which is characteristic in many of devitrified glasses in the CaO–MgO–SiO2 system [30]. In contrast, a small amount of amorphous phase seems evident following calcination of solid state samples and can be attributed to unreacted precursor particles.

The presence of $SiO_2$, predominantly in the form of cristobalite phase, in the sintered samples indicates the crystallisation of the aforementioned amorphous silica rich glass that was present to varying degrees following calcination. The phase compositions shown in Fig. 1 do not include any residual amorphous phases that may be present after sintering. However, the absence of an observable amorphous halo in XRD patterns of sintered materials indicates that any amorphous content in these materials is unlikely to exceed 1 wt%, and can be assumed to be limited to silica rich interface regions in a polycrystalline material [31].

The synthesis approach applied shows a clear influence on the sequence and extent of phase crystallisation. For example, MgO and CaO form as intermediates phases during thermal processing of Pechini-derived materials, but did not appear to form following calcination of the precipitated silicates, which followed a phase formation pathway more characteristic of glass devitrification, indicating that an intermediate glassy phase mediates the observed crystallisation and densification behaviour seen in CP materials.

In co-precipitated magnesium-rich samples, a magnesium deficiency is observed relative to the target crystalline phase. This indicates that in the method applied here, solute $Mg^{2+}$ ions remain for precipitation at silica-lean stoichiometries, which can then be lost in the removal of the supernatant. The precipitation of Mg-containing species is highly dependent on the precursor materials used, which has been demonstrated in detail by other authors [32]. Other deviations in stoichiometry may further be produced as the result of cation volatilization. In the quantification of crystalline phases in Fig. 1 a silica deficiency can be observed for willemite and enstatite (presence of unreacted ZnO and forsterite, respectively) samples produced with the standard solid state method. Elemental mapping indicates that the missing Si is present in the residual amorphous phase.

### 3.2. Microstructure

Particle size distributions (PSD) for powders in both unfired and calcined forms are presented in Fig. 2. As the measured PSDs were similar for all compositions in each production route, the values were averaged for the sake of clarity. Powder products from SS and P synthesis show comparable distributions with a bimodal character after calcination and milling. The ceramic products derived from the CP route show a very broad unimodal distribution. Interestingly, calcined CP powders respond well to milling after calcination as the averaged d50 value decreased from 21.6 μm to 1.7 μm, significantly smaller than those values for SS and P.

A microstructural overview for all sintered samples is presented in the supporting information (Fig. S3). In the following, characteristic microstructural differences of pellets produced with the different synthesis methods will be presented exemplarily with samples with nominal compositions of $Mg_{1.8}Zn_{0.2}SiO_4$ and $Zn_2SiO_4$. Additionally, an elemental EDS mapping of $Mg_{1.8}Zn_{0.2}SiO_4$ samples will be shown exemplarily.

In Fig. 3, the microstructures of sintered pellets of $Mg_{1.8}Zn_{0.2}SiO_4$

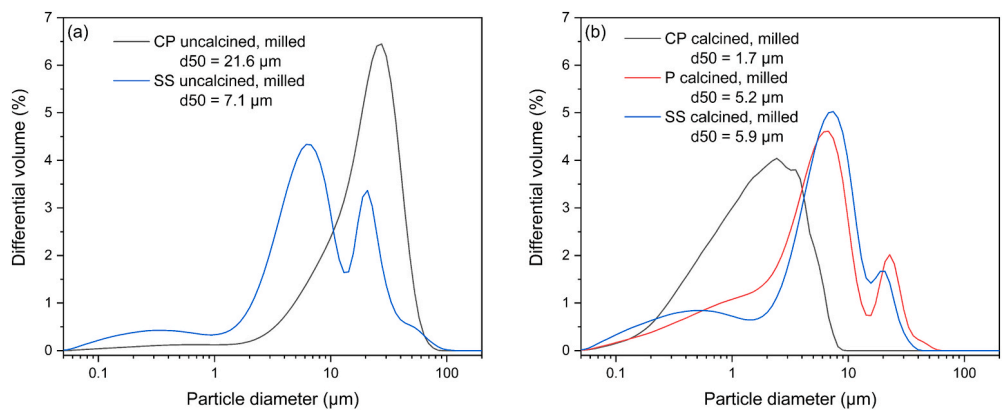

**Fig. 2.** Particle size distributions for uncalcined silicate powders (a) and calcined powders (b). The values inside the CP, P and SS series were comparable for each of the 8 compositions, an average is displayed for the sake of clarity.

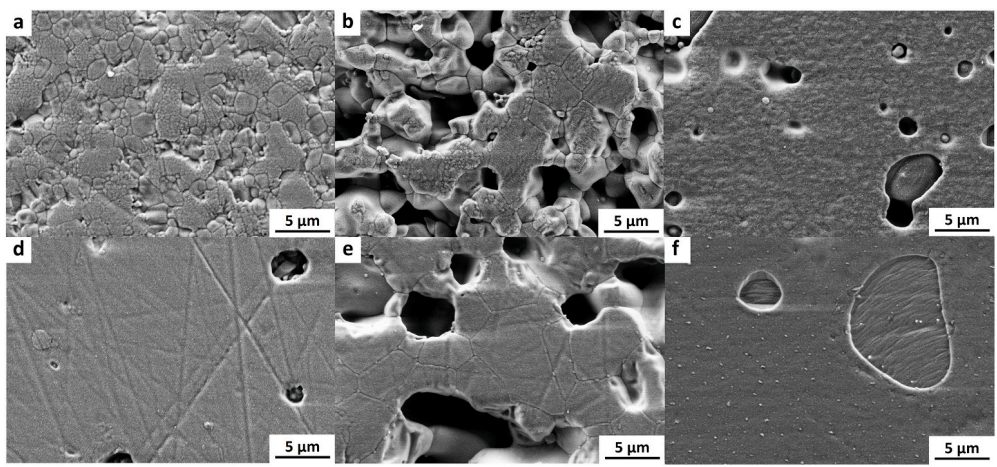

**Fig. 3.** Microstructure of sintered pellets of $Mg_{1.8}Zn_{0.2}SiO_4$ (a-CP, b-P and c-SS), and $Zn_2SiO_4$ (d-CP, e-P and f-SS) nominal composition.

and $Zn_2SiO_4$ (CP, P and SS) are depicted, showing significant differences. While CP and SS $Mg_{1.8}Zn_{0.2}SiO_4$ (a and c) and $Zn_2SiO_4$ (d and f) powders are sintered into dense pellets with open porosity values < 1% (Table 1), treating P powders with the same temperature profile will not suffice to close residual pores and results in a poorly sintered microstructure with high porosity (Fig. 3b and e). When comparing Fig. 4a and c, it is striking that the SS derived pellets appear to consist of a homogeneous matrix with inclusions of unreacted material, while sintered CP materials show mainly discrete grains. This observation can be confirmed by elemental analysis with EDX (Fig. 4c), showing that comparably large particles of unreacted binary oxide (Mg,Zn)O remain in the sintered structure of the SS sample. Besides that, the matrix-like volume of this pellet is chemically homogeneous. In contrast, Mg, Si and Zn clearly show segregation phenomena in the $Mg_{1.8}Zn_{0.2}SiO_4$ sample prepared via co-precipitation (Fig. 4a), in which the areas with a higher Mg ratio can be assigned to Forsterite and the ones with higher Si to clinoenstatite (Fig. 1a). On the other hand, the sintered P pellet (Fig. 4b) is entirely homogeneous, showing no inclusions or segregations sites. The bright spots in the picture are representations of porosity and show no elemental signals.

### 3.3. Thermal analysis

The behaviour of the different sample powders upon heating was studied exemplarily on $Mg_2SiO_4$ and $Zn_2SiO_4$ compositions and is presented in Fig. 5 and Fig. 6. When heating uncalcined CP powders (Fig. 5a and d), main weight loss occurred in the temperature range 100–200 °C by, endothermic vaporization of water. No more mass loss occurs at higher temperatures. The temperature behaviour of gels derived from the Pechini method (Fig. 5b and e) is characterised by a multistage combustion sequence with relatively sharp weight loss maxima at around 150 °C, 200 °C, 300 °C and 500 °C, which can be assigned to the volatilization of residual solvent, decomposition of nitrates and formation of $NO_x$ and $N_2$ species, loss of organics by pyrolysis and release of $CO_2$, respectively [33]. In our Pechini synthesis, we used a molar ratio of metals, citric acid and ethylene glycol of 1:6:6, which is considerably higher than usually reported for the synthesis of oxide materials [34], but showed the best results in our initial trials. Because of this high ratio, the great mass loss during heating is not surprising here. In the case of the SS samples, the heating of milled starting powder for $Mg_2SiO_4$ to 800 °C results in two distinct mass loss events between $100-150$ °C and $400-450$ °C, respectively. The total loss of 27.4% can be attributed to water release, as it is known that MgO is a very hygroscopic compound [35]. On the other hand, heating powder mixtures of ZnO and $SiO_2$ to 800 °C is uneventful and exhibits minimal mass loss (Fig. 5f).

Upon heating powders to 1400 °C after calcination, milling and drying shows (Fig. 6a) a sharp mass loss event at around 400–450 °C in samples containing MgO after calcination (P and SS $Mg_2SiO_4$, Fig. 6b and c), which can be attributed to the removal of chemisorbed $H_2O$ species in MgO [36,37]. The mass loss in the CP $Mg_2SiO_4$ sample, not containing MgO, is limited to evaporating surface water at around 200 °C (Fig. 6a). Crystallisation and new phase formation commences at around 850 °C in all $Mg_2SiO_4$ samples and is characterised by the release of latent heat of solidification seen as a broad exothermic event between

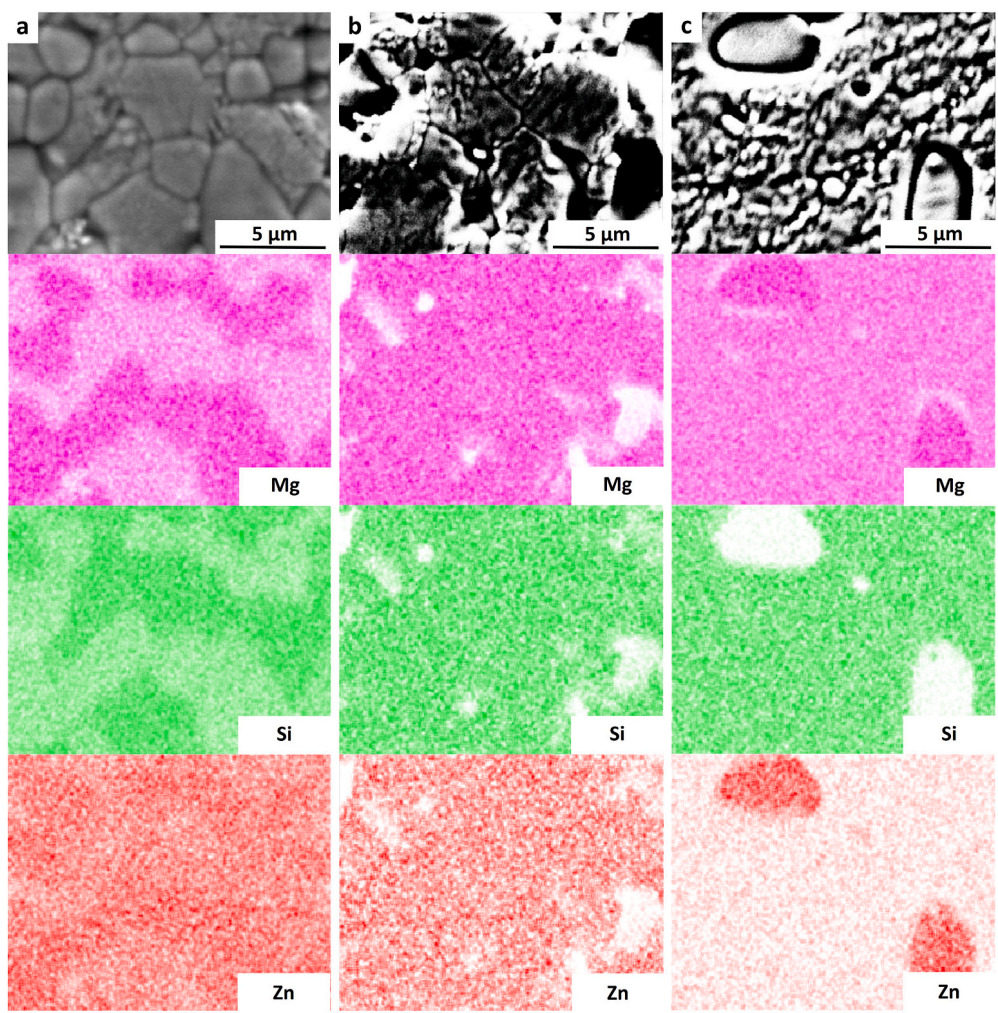

**Fig. 4.** Elemental mapping of Mg, Si and Zn for sintered zinc substituted forsterite $Mg_{1.8}Zn_{0.2}SiO_4$ synthesised by precipitation, (column a), the Pechini method (column b) and the solid state method (column c).

850 and 1300 °C. Very limited or no mass loss can be observed when heating calcined powders of $Zn_2SiO_4$ (Fig. 6d, e and f). While the onset of crystallisation for the CP sample (Fig. 6d) can be observed at around 800 °C, it is significantly lower in temperature (roughly 600 °C) and progresses over a broader temperature range for the p and SS samples (Fig. 6e and f).

The sintering behaviour, derived from HSM measurements of the samples with nominal $Zn_2SiO_4$ composition, is depicted in Fig. 7. Here, the term form factor refers to changes from the initial shape (1) of the pressed sample, e.g., rounding of edges. The size of the samples' shadows cast on the camera is referred to as area and is also 1 in the starting state. When the sample area decreases while the form factor remains unchanged (isotropic shrinkage), sintering of the sample takes place. This is the case for all 3 samples in Fig. 7 in the temperature range 1100–1350 °C. However, significant differences can be observed in the rates with which the area decrease takes place, indicating different rates of sintering between materials produced with the three methods. The greatest degree of densification is achieved by the co-precipitated sample. Interestingly, the solid-state sample did not compact any further once the dwelling temperature (1350 °C) was reached and held for 30 min. The sol-gel derived sample showed the poorest sintering behaviour. The diametral shrinkage values are 28.3% for the CP sample, 13.3% for the P sample and 25.0% for the SS sample.

## 4. Discussion

### 4.1. Phase formation

Among oxide ceramics, silicates are known to exhibit convoluted phase formation behaviour, which has been studied extensively in the realms of earth sciences. Here we focus on silicates as functional ceramic materials and examine the formation of desired phases. The phase formation behaviour in silicates is uniquely complex and is often hard to predict on the basis of composition alone. This is because solid state reactions between intermediate silicate phases do not occur with appreciable kinetics even at temperatures close to the melting point of one of the phases, resulting in highly stable non-equilibrium phase mixtures in sintered products. Consequently, obtaining a targeted phase and composition and microstructure in silicate ceramics, which is often important towards the design of highly performing materials, requires the informed selection of appropriate synthesis and processing methods.

This study is motivated by the utility of soft-chemistry methods towards the design of functional silicate ceramics. It is well known that such methods can produce non-equilibrium phases and diverse microstructures in oxide system in general, and understanding the limitations and benefits of the available approaches for the synthesis of silicate ceramics, in particular, is of value towards improved materials design and processing in this versatile and promising group of ceramics materials. The most widely applied method of solid-state reactions between oxide precursors exhibits some notable shortcomings towards the fabrication of functional silicate ceramics. In the rapidly growing silicate ceramics field, recent studies that have utilised relatively conventional solid state methods reveal notable limitations. Very few studies have reported the fabrication of synthetic sorosilicate in dense bulk and pure phase form. Moreover, those studies in which such ceramic phases have been synthesised often rely on high temperature processes, involving the formation of a melt, and the subsequent presence of significant amorphous silicate-based glass phases. For the obtainment of polycrystalline silicate ceramics without any $SiO_2$ dominated glasses, synthesis techniques must be designed with target microstructure and assemblage in mind.

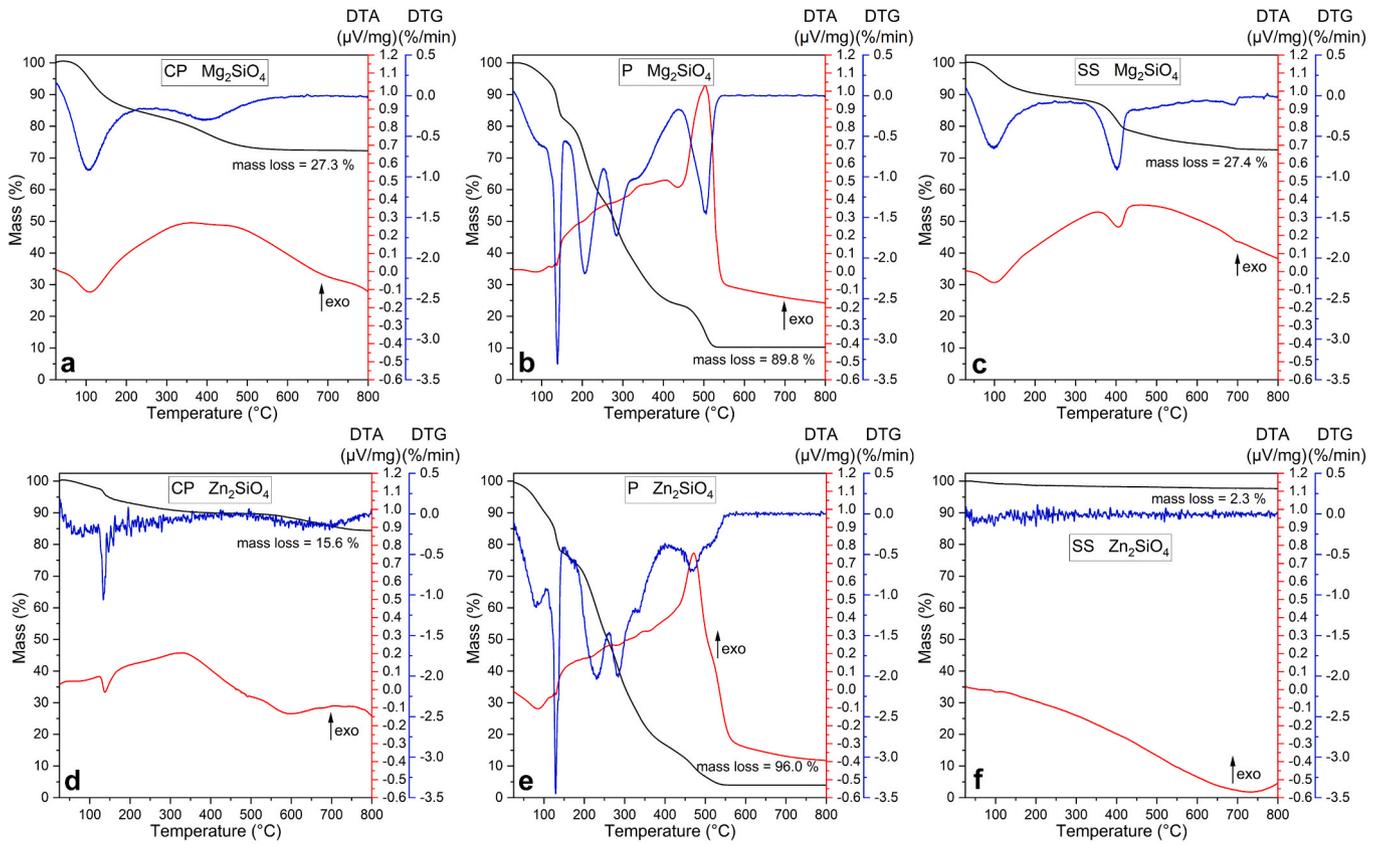

**Fig. 5.** Thermal analysis data for uncalcined samples heated from 20 to 800 °C.

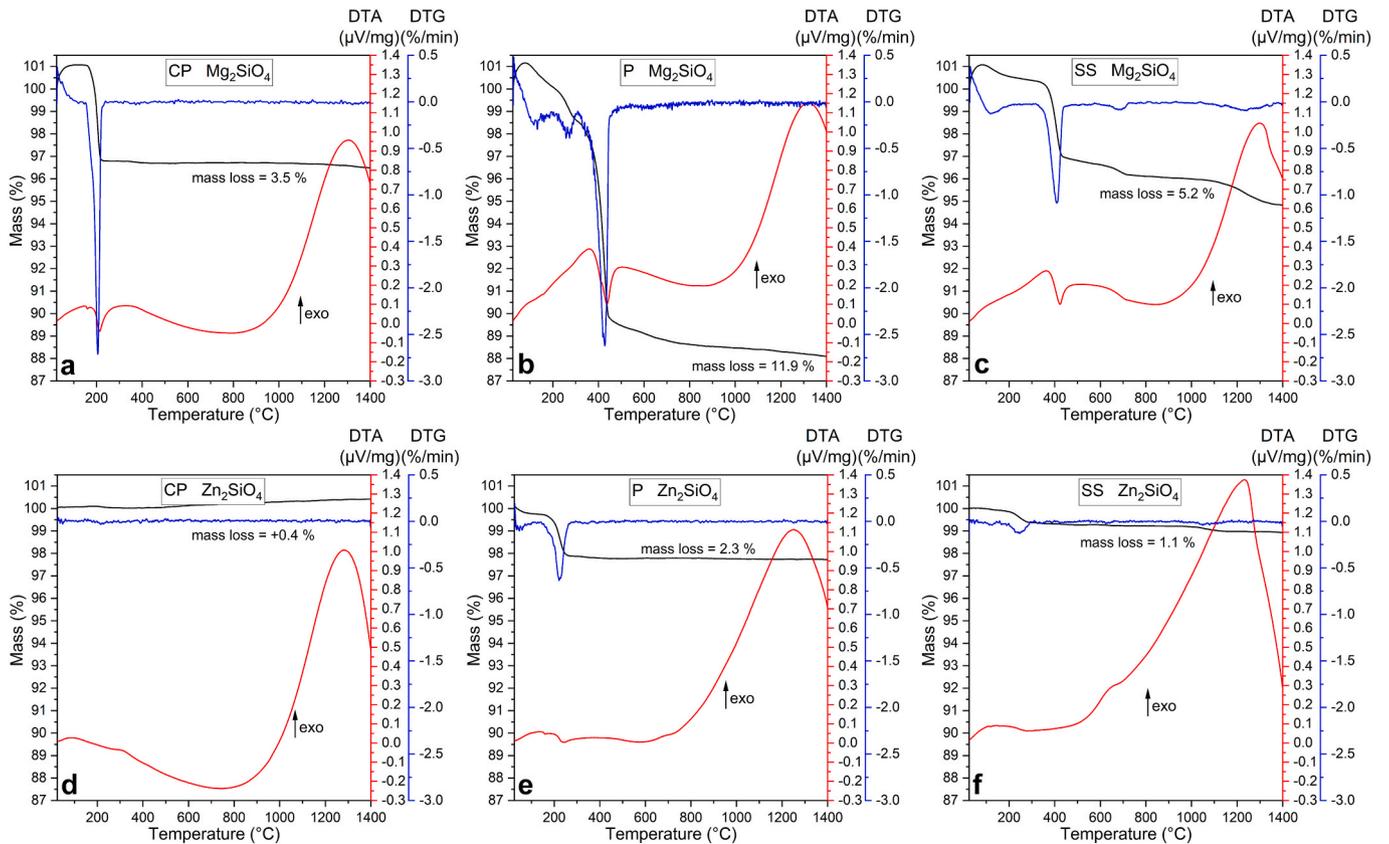

**Fig. 6.** Thermal analysis date for calcined, milled and dried powders heated from 20 to 1400 °C.

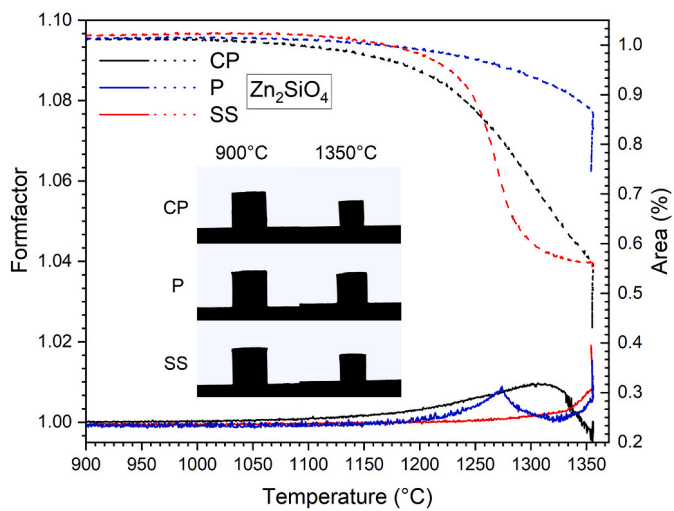

**Fig. 7.** Results of the HSM measurements represented in changes in form factor (solid lines) and area (dashed lines) of the pressed sample cylinders. The development of the sample shapes in the temperature range from 900 to 1355 °C is shown in the insert.

From Fig. 1, it is quite clear that in our studies employing simple variations of co-precipitation, sol-gel and standard solid-state synthesis types, the Pechini method yields the best results in terms of desired silicate phase formation. Although the akermanite composition shows significant amounts of secondary phases, it would seem that the steric entrapment afforded by the chelation-polymerisation process in the Pechini method allows for homogenous multi-cation silicates to form in phase-pure products. Although, due to reaction kinetics, the combusted Pechini resin does not directly form a single crystalline phase straight from calcination, short diffusion paths in such materials mean that a well equilibrated final phase assemblage after sintering is readily achieved through this method.

In the fabrication of silicate ceramics via a precipitation route, several advantages exist. The completion of phase formation occurs more rapidly in materials synthesised in this way. This may be the result of the initial short-range order formed in a precipitate, which is most often in the form of an amorphous hydroxide. The coordination of cations in a formed precipitate is strongly dependent on precipitation conditions, and thus the selection of precursors (e.g. differences in solubility of nitrates and chlorides), pH and solvents will all produce profound effects on the composition and homogeneity of an initial precipitate. As seen here, the removal of a supernatant, which may still contain precursor cations, can result in stoichiometric distortion of the product. In this case, the loss of magnesium ions encumbered the obtainment of phase pure forsterite. The results of the present work underline the need for precipitation based soft chemistry routes to be specifically adapted and optimised towards specific silicate families and compositions. As an example, the obtainment of enstatite by precipitation in phase pure form has been reported by Ref. [38], where nitrate precursors were reacted with sodium metasilicate. Slight variations in cation occupancy are likely the reason why in our work a metastable protoenstatite polymorph was produced via a precipitation route, while a clinoenstatite phase, which is metastable, was obtained when a different precipitation reaction pathway was employed [38]. It should be noted that from the results here, it is clearly seen that zinc has a profound effect on phase formation in polymorphic enstatite, stabilising the low temperature polymorph orthoenstatite. In solid state methods, diffusion distances tend to be longer, with homogeneity depending primarily on the mixing and comminution of precursor powders. Despite milling and consolidation of powders, solid state materials produced here tended to exhibit unreacted binary oxides following sintering, either MgO, ZnO or $SiO_2$. This is the result of particles in the sintered compact remaining isolated from contact with other constituents.

Silicates are a special class of oxides, comprising silica tetrahedra as their fundamental building block. In all three synthesis approaches studied here, phase formation involves a step of metal cations migrating into a corner-sharing $SiO_2$ network, breaking up the initial tetrahedral connectivity of the network, which provides a static and open framework into which these cations can diffuse comparably easily [14,39]. This has been confirmed for ceramics derived from the co-precipitation [16], the sol-gel [13] and the solid-state method [12]. As indicated by the variation of phase assemblages between samples prepared by different methods, it would be logical to assume that synthesis affects primarily the homogeneity and diffusion pathways of cations into the host silica network.

In the following, we will assess the differences in phase formation mechanisms for the three routes on the example of forsterite ($Mg_2SiO_4$). From a precursor containing Mg and $SiO_2$, the formation of Enstatite typically precedes that of Forsterite, by migration of $Mg^{2+}$ into $SiO_2$ [39, 40]. If in some synthesis method explored here, this would be different, significant differences in the DTA curves of Fig. 6a–c would be observable because it is reported that the crystallisation of enstatite is very exothermic, while the of forsterite is rather drawn out without clear exothermic events [41]. Since all three DTA curves in Fig. 6 are similar, it can be expected that enstatite (namely orthoenstatite) is the first crystalline Mg–Si compound in nominal $Mg_2SiO_4$ samples for CP, P and SS. For the conversion of $Mg_2Si_2O_6$ to $Mg_2SiO_4$ further diffusion of $Mg^{2+}$ is required. The reason that this reaction is faster for P is probably the state in which MgO and $SiO_2$ are present. While in SS and CP those phases exist as distinct particles before temperature treatment, MgO and $SiO_2$ are known to form more chemically uniform xerogels or even core-shell structures if derived from sol-gel syntheses [42]. The reactivity of this homogenous mixture is very likely to be higher than that for particulate MgO and $SiO_2$ where the reaction happens mainly at grain-grain contacts with highly localised diffusion pathways. This way, we can also explain the very homogenous elemental distributions in Fig. 4b for sample P, while CP and SS are not fully reacted.

The formation of willemite ceramics ($Zn_2SiO_4$) seems to progress readily for all syntheses, although the reaction is not fully completed in the solid-state samples (Fig. 1b). Since there is no intermediate phase in the reaction 2 ZnO + $SiO_2$ → $Zn_2SiO_4$ [43,44], as is the case for the forsterite phase ($Mg_2SiO_4$) and the fact that Zn diffusion rates are higher than those for Mg in enstatite and forsterite [14], this compound is readily synthesisable with any of the 3 methods. Further, synthesis of enstatite samples ($Mg_2Si_2O_6$, Fig. 1c) with the CP method is more straightforward than forsterite, as only secondary phases in CP samples for $Mg_2Si_2O_6$ and $Mg_{1.8}Zn_{0.2}Si_2O_6$ are considerably smaller than for $Mg_2SiO_4$ and $Mg_{1.8}Zn_{0.2}SiO_4$. As no intermediate reaction is necessary for the conversion of MgO and $SiO_2$ into $Mg_2Si_2O_6$, the chemical equilibrium is reached faster here, analogous to willemite samples. Interestingly, the opposite is true for SS samples, where forsterite samples show better results in terms of phase compositions than the one for enstatite. Clinoenstatite is the most common phase in samples $Mg_2Si_2O_6$ and $Mg_{1.8}Zn_{0.2}Si_2O_6$, which is to be expected as cooling enstatite samples from temperatures well above 1000 °C typically leads to the conversion of the high temperature polymorph protoenstatite to clinoenstatite [45,46]. On the other hand, CP $Mg_2Si_2O_6$ is the only sample containing protoenstatite. The simultaneous presence of clinoenstatite here indicates that the conversion of PE to CE is retarded in enstatite prepared with the CP method. This behaviour is particularly noteworthy as this conversion is reported to be of martensitic and diffusionless character and usually happens rather quickly [45]. As discussed earlier, for the synthesis of the complex calcic sorosilicate -akermanite (Fig. 1d), $Ca_2MgSi_2O_7$ is present as major phase only in materials synthesised with by the Pechini-type method. For samples SS $Ca_2MgSi_2O_7$ and $Ca_2Mg_{0.9}Zn_{0.1}Si_2O_7$ the ceramic product comprises a mixture of merwinite and monticellite intermediates, indicating that the reaction is not completed after the chosen firing profile. The formation of diopside ($CaMgSi_2O_6$) can be observed as the main phase in CP samples of $Ca_2MgSi_2O_7$ and $Ca_2Mg_{0.9}Zn_{0.1}Si_2O_7$, indicating either a deficiency in Ca or its presence in X-ray amorphous form. The formation of diopside is known to occur very readily in co-precipitation derived materials as well as in crystallisation from glasses [30,47,48], suggesting that the favourable formation kinetics of this phase needs to be considered in the design of calcic sorosilicates. As indicated here, in the synthesis of sorosilicate ceramics the use steric entrapment based.

### 4.2. Effect of Zn and Mg substitution

Ionic substitution is a popular tool to tune or optimise relevant material properties by taking advantage of the fact that most minerals exhibit a considerable range in chemical composition. In the simplest case of cationic or anionic substitutions, an ion A will be replaced by an ion B. Goldschmidt explained fundamentally that in this case, ions that are closest in radius and charge are the easiest to substitute for each other [17]. Prominent examples of ionic replacements for the tuning of material parameters are found in almost every materials science direction. In the ongoing search for materials with ever improved dielectric properties for high frequency wireless applications, cation substitutions are often introduced in the host material to reduce dielectric loss [49–51], achieve near zero temperature factor of resonant frequency [52,53], reduce secondary phase formation [54], improve densification and adjust its permittivity [1]. Bioactive glasses and ceramics are routinely substituted with a broad range of cations to harness their profound effects on the material properties. Examples include Sr (enhancement of metabolic activity in osteoblasts [55]), Ag (antibacterial activity [56]), Zn (stimulates osteogenic activity, bone in-growth and healing [57,58]) or Cu (improved angiogenesis [59]). Exchanging ions of different sizes can lead to the stabilisation of a metastable polymorph of a material, allowing its use at ambient conditions. The prime example of this approach is the stabilisation of tetragonal and cubic zirconia by the addition of yttria and the accompanying substitution of Zr cations with Y in the lattice [60]. Besides numerous other utilisation examples in i.a. sensors, catalysis and electrochemical devices, ionic substitutions are increasingly employed for the discovery of novel compounds, postulating the existence of an unknown compound by replacing ions in known materials by a chemically similar ion [61].

Silicates exhibit broad ranges of solid-solution formation, arising from the pliability of the Si–O–Si bond angle in corner-shared $SiO_4$ tetrahedra, ranging from $120°$ to $180°$ [62]. This angle flexibility allows silicate structures to withstand structural distortions, evoked by the introduction of cations of varying sizes.

The biggest impact of substituting Zn can be observed in the phase formation of enstatite polymorphs. In this study, following sintering, the orthoenstatite polymorph is only present in CP and SS $Mg_{1.8}Zn_{0.2}Si_2O_6$ (Fig. 1), both Zn-substituted samples. Usually, to obtain OE, a very slow cooling regime is required as the conversions PE→OE and CE→OE are sluggish order-disorder types that require a lot of time [45]. Since we can not observe the presence of OE in unmodified $Mg_2Si_2O_6$ we can conclude here, that Zn enhances the transformation of high temperature polymorphs PE and CE into OE. From the EDX analysis (Fig. 4), two further conclusions can be drawn: (i) in samples containing $Mg_2SiO_4$ as well as $Mg_2Si_2O_6$ (sample CP $Mg_{1.8}Zn_{0.2}SiO_4$, Fig. 4a), Zn is preferentially found in $Mg_2SiO_4$ indicating that Zn incorporation into forsterite is energetically more favourable than into enstatite. And ii) when using sufficiently large MgO precursors as in sample SS $Mg_{1.8}Zn_{0.2}SiO_4$, ZnO will firstly dissolve into the rocksalt structure of periclase (MgO) to form $Mg_{1-x}Zn_xO$, which forms a solid solution [63]. This is evident in Fig. 4c, which shows that Zn is almost exclusively present in areas with high amounts of Mg and no Si (MgO). Due to the wide solid-solution regimes between willemite and forsterite the substitution of Mg in willemite, is readily achievable without the formation of additional secondary phases through all synthesis methods. As expected, the change in chemistry produces a fluxing effect, with Mg substituted willemite showing enhanced sinterability. The substitution of Mg with Zn was further found to be a way to reduce the formation of the secondary enstatite in co-precipitation of a target forsterite phase. As the soft-chemistry synthesis of forsterite has several promising applications in components that rely on density and improved fracture toughness [64,65], improving the formation and sintering of this phase is a useful objective. The enhancement of forsterite formation is similar to that which occurs when Ti is substituted for Mg in forsterite, except that this substitution is achievable to a greater extent, as high levels of Ti produce other secondary phases [54]. It is unclear whether the enhanced formation of forsterite in substituted co-precipitated compositions can be attributed solely to a fluxing effect and a higher homologous temperature, or whether the phase purity can be attributed to a decreased formation energy of the substituted lattice or to the formation of particularly reactive intermediate phases (such as ZnO). This could motivate more elaborate studies, including ab-initio calculations. From our results into Zn substitution in akermanite, it would appear that a simple fluxing effect is not the only factor at play here, as a Zn substituted showed poorer phase formation behaviour, despite the higher homologous temperature.

### 4.3. Sintering behaviour

We have examined densification in both forsterite and willemite prepared by different methods. Due in part to their high level of symmetry, both of these orthosilicate phases are interesting as high-performance low loss dielectric materials [66]. In the fabrication of dielectric ceramics in useful components, lowering the densification temperature is useful to accommodate co-firing processes, e.g. for LTCC materials. Additionally, improving the densification of material will contribute to improved dielectric loss characteristics. Further, in the field of bioceramics, densification is also important in order to achieve good fracture toughness in polycrystalline materials, with relevance for applications in bone-tissue engineering scaffolds.

Based on the findings seen here in TGA as well as hot stage microscopy, silicate ceramics processed by co-precipitation exhibit superior densification behaviour, due possibly to a high fraction of densely packed crystallites of metastable or reactive intermediate phases, including considerable amorphous fraction. Where formation kinetics are favourable, this rapidly produces a denser ceramic of the target phase. It would seem that such a precipitation route should be applied where the target application is served by a dense polycrystalline silicate that exhibits good formation kinetics. Certain studies have reported that the inclusion of substitutional cations can be designed to minimize the tendency of secondary phase formation, and thus the adjustment of composition can be used in the future to enable the precipitation of homogenous ceramics for materials that would otherwise exhibit unfavourable phase formation kinetics. In particular, the substitution of $Si^{4+}$ by $Ti^{4+}$ in forsterite synthesis by a solid state method was shown to reduce the formation of the undesired enstatite phase [54], which was crystallised significantly in the co-precipitation derived materials studied here. The results found here, as discussed, showed a similar enhancement of orthosilicate formation when $Mg^{2+}$ was substituted with $Zn^{2+}$. Future work should examine the rational design of cation substitution in order to enhance both phase purity and sintering behaviour of co-precipitated orthosilicates, with forsterite, in particular, being of interest for a multitude of potential applications.

Precipitated powders and solid state precursors exhibit very low mass loss during calcination as indicated by TGA. In contrast, materials formed by the Pechini-type sol-gel method, contain a high organic content due to the chelation and polymerisation groups that are lost through combustion. This results in a mass loss during calcination of approximately 90% and consequently a highly porous calcined product. The subsequent densification of Pechini derived pellets proceeds more slowly as indicated by Fig. 7.

### 4.4. Processing – performance relationships

The studies here confirm that co-precipitated materials offer attractive pathways towards achieving desired combinations of density and phase assemblage in functional silicates. Several existing studies have applied precipitation methods for the fabrication of ceramics for biomedical, phosphor, dielectric or magnetic applications [67–70]. Based on the observations found in this study, as well as those which have been reported from other forays into precipitation of functional silicates [16,38,47,48,71] certain phases form very readily by precipitation, such as diopside, enstatite and willemite. However, as seen here and as reported elsewhere, other phases are not amenable for single phase formation through co-precipitation pathways, and large amounts of secondary phases are present. The observation here that zinc substitution enhances forsterite formation in co-precipitation methods motivates further studies to evaluate approaches to adjust a silicate's composition in order to simultaneously enhance phase purity of a precipitated product, enhance sintering and densification and further adjust functional properties.

Where high control of phase homogeneity is required, with appropriate sintering parameters, the versatile phase assemblages afforded by steric immobilisation based synthesis can too be formed in dense microstructures.

As we have shown here, the choice of synthesis and thermal processing methods for silicate ceramics profoundly affects their characteristics, and the design of such methods needs to be made with a target composition and application in mind. The crystallography, microstructure, porosity and phase assemblage of applied ceramic materials govern their performance in various applications. Based on the results found here, each of the studied methods has its advantages and disadvantages with respect to applied ceramic materials.

A precipitation route is advantageous where:

1. A scalable process can be implemented to produce relatively phase pure materials of diopside or willemite
2. Secondary phases are not deleterious to performance
3. Enhanced densification is beneficial

The use of steric-immobilisation is advantageous where:

1. Homogeneous cation distribution is required
2. Formation kinetics are unfavourable, and precipitation methods are ineffective
3. High porosity is not detrimental or is even helpful
4. High levels of defects and surface sites are tolerable

In bioceramics intended towards bone tissue engineering applications, osseointegrative bioactivity alongside biodegradability are often sought after [72].The use of crystalline silicates, as opposed to bioglasses, may be advantageous in terms of strength and porosity. For such applications, distorted nanocrystallites with high surface area often confer enhanced dissolution kinetics, enabling the inter-growth of bone into resorbable polycrystalline ceramic scaffolds. The strength of bioceramics is inversely related to their porosity. In non-load bearing scaffolds a highly porous nanocrystalline microstructure such as that produced through the use of soft-chemistry based methods may facilitate an optimised balance of strength and bioactivity. The results of the present work might motivate the utilisation of porous silicate materials produced via the Pechini method as ceramic membranes. Here, silicates could be a low-cost production alternative with further attractive attributes such as high-temperature stability and good chemical resistance necessary for applications such as water purification [73].

In silicate ceramics for dielectric applications, high density and phase purity are desirable. The ability to process such ceramics at lower temperatures is of further value in allowing the incorporation of silicate dielectrics into co-fired devices.

## 5. Conclusion

In this work, we comprehensively examined soft-chemistry synthesis approaches in the form of sol-gel (Pechini) and co-precipitation methods and compared them with a standard solid-state approach to assess the suitability of the different techniques towards the fabrication of silicate ceramics. The results were interpreted in terms of morphology, phase formation, and sintering characteristics. Based on our findings, we can here formulate the following guidelines for the synthesis of silicate ceramics:

1. Target stoichiometry in complex silicate ceramics can be most easily attained by employing the modified sol-gel method developed by Pechini [24], in which steric cation entrapment and homogeneity on a molecular level prevent phase segregation and allow for short diffusion pathways. The method is very robust and repeatable; powders produced with this method, however, have poor sinterability and require higher sintering temperatures compared to co-precipitated or solid-state derived powders.
2. Product powders formed via co-precipitation have excellent sintering characteristics and allow for rapid densification comparable to solid-state methods. To attain the desired silicate stoichiometry with a co-precipitation method, however, is not as straightforward and robust as the Pechini method and requires fine tuning of all synthesis parameters, as the precipitation of metal hydroxides is extremely sensitive to a multitude of factors.
3. For solid-state synthesis, equilibrium phase formation is very slow compared to soft-chemistry approaches. Since the densification of solid-state derived powders is comparably quick, calcination times of the initial powder mixture should be increased compared to soft-chemistry methods to ensure complete reaction before sintering. The main merit of the solid-state method is its simplistic nature, low cost, and repeatability, but in terms of the product characteristics examined here, it does not offer clear advantages compared to soft-chemistry approaches.

### CRediT author statement

**F. Kamutzki**: Conceptualization, Methodology, Investigation, Data Curation, Visualization, Writing - Original Draft Preparation. **M. F. Bekheet**: Data Curation, Validation, Writing - Review & Editing. **S. Schneider**: Investigation, Data Curation. **A. Gurlo**: Supervision, Writing - Review & Editing, Project administration, Funding acquisition, Resources. **D.A.H. Hanaor**: Conceptualization, Methodology, Supervision, Writing - Original Draft Preparation, Writing - Review & Editing, Project administration.

### Funding

This research did not receive any specific grant from funding agencies in the public, commercial, or not-for-profit sectors.

## Declaration of competing interests

The authors declare that they have no known competing financial interests or personal relationships that could have appeared to influence the work reported in this paper.

## Acknowledgements

We would like to thank Delf Kober from Technische Universität Berlin for thermal analysis. Further, we express our gratitude towards Dagmar Nicolaides from Bundesanstalt für Materialforschung-und prüfung (BAM) for HSM measurements.

## Appendix A. Supplementary data

Supplementary data to this article can be found online at https://doi.org/10.1016/j.oceram.2022.100241.